\documentclass[preprint,sort&compress,review]{elsarticle}
\usepackage{tikz}
\usepackage{amsmath}
\usepackage{amssymb}
\usepackage{pgfplots}
\usepackage{amsthm}
\usepackage{subfigure}
\usepackage{tabularx}
\usepackage{multirow}
\usepackage{hyperref}
\usepackage{mdframed}
\usepackage{breqn}
\usepackage{array,booktabs}
\usepackage{tablefootnote}
\usepackage{standalone}
\newcommand{\mbf}[1]{\mathbf{#1}}
\newcommand{\lrc}[1]{\left\{#1\right\}}
\newcommand{\lrs}[1]{\left[#1\right]}

\newcommand{\figDir}{./SetA1/p1/fig/TikZ}

\newenvironment{eqnArrAligned}
{\begin{eqnarray}\begin{aligned}}
{\end{aligned} \end{eqnarray}}

\newenvironment{eqnArrAligned*}
{\begin{eqnarray*}\begin{aligned}}
{\end{aligned} \end{eqnarray*}}

\newtheorem{lemma}{Lemma}

\begin{document}

\begin{frontmatter}

\title{A Risk Minimization Framework for Channel Estimation in OFDM Systems \tnoteref{ack}}

\author[iisc1]{Karthik Upadhya \corref{cor1}} \ead{karthik.upadhya@gmail.com} 
\author[iisc2]{Chandra~Sekhar~Seelamantula} \ead{chandra.sekhar@ee.iisc.ernet.in}
\author[iisc1]{K.V.S.~Hari} \ead{hari@ece.iisc.ernet.in}

\address[iisc1]{Department of Electrical Communication Engineering}
\address[iisc2]{Department of Electrical Engineering, Indian Institute of Science, Bangalore - 560 012, India}
\cortext[cor1]{Corresponding author. Phone No. : +91-7353585508}

\tnotetext[ack]{The financial support of the DST, India and of the EPSRC, UK under the auspices of the India-UK Advanced Technology Centre (IUATC) is gratefully acknowledged.}
% [TBD] Credit for project

\begin{abstract}
We address the problem of channel estimation for cyclic-prefix (CP) Orthogonal Frequency Division Multiplexing (OFDM) systems. We model the channel as a vector of unknown \emph{deterministic} constants and hence, do not require prior knowledge of the channel statistics. Since the mean-square error (MSE) is not computable in practice, in such a scenario, we propose a novel technique using Stein's lemma to obtain an unbiased estimate of the mean-square error, namely the Stein's unbiased risk estimate (SURE). We obtain an estimate of the channel from noisy observations using linear and nonlinear denoising functions, whose parameters are chosen to minimize SURE. Based on computer simulations, we show that using SURE-based channel estimate in equalization offers an improvement in signal-to-noise ratio of around $ 2.25 $ dB over the maximum-likelihood channel estimate, in practical channel scenarios, without assuming prior knowledge of channel statistics.
\end{abstract}


\end{frontmatter}

\section{Introduction}
%\subsection{Background}
%** Setting, Background and Importance of the problem ** \\
Orthogonal Frequency Division Multiplexing (OFDM) has become a popular modulation scheme for various mobile \cite{3gppPhy}, wireless networking, and digital television standards \cite{dvb-t,dvb-t2}. One of the primary advantages of OFDM over single-carrier modulation schemes is its superior performance in multipath channels. Estimating the value of the channel frequency response (CFR) at each sub-carrier accurately is crucial for recovering the transmitted data from the detrimental effects of the channel. To aid in channel estimation, typical OFDM systems reserve a part of the sub-carriers for pilots. These pilots can either be fractions of the sub-carriers in every OFDM symbol or be aggregated into one OFDM symbol, called the preamble. Preamble symbols are used for obtaining initial channel estimates, which can be further refined using pilots interleaved within subsequent data symbols. 

%** Related Work **
\subsection{Related work on channel estimation for OFDM systems}
In a survey paper, Ozdemir and Arslan \cite{Arslan2007} considered channel estimation based on the maximum-likelihood (ML) criterion and the linear minimum mean-square error (LMMSE) criterion using a preamble OFDM symbol. In the ML approach, the channel is assumed to be a vector of unknown deterministic constants. When the noise is additive and white-Gaussian, the channel estimation is obtained by minimizing the Euclidean distance between the observation and the parameter to be estimated. At low signal-to-noise ratio (SNR), the algorithm fits the estimated parameter to the noise in the observation and hence, the ML approach has limited performance. On the other hand, Bayesian techniques such as the LMMSE estimator \cite{Mengali2001} outperform the ML estimate, but require a priori knowledge of the autocorrelation matrix of the CFR, which may not always be available in practice. In addition, the LMMSE channel estimator has a high complexity, because it requires an inversion of the channel autocorrelation matrix. Considerable efforts have been devoted to reducing the complexity as well as to render the estimator robust to inaccuracies in the knowledge of the channel statistics. In particular, van de Beek et al. \cite{vanDeBeek1995} obtained a reduction in the size of the matrix to be inverted assuming prior knowledge of the length of the channel impulse response (CIR). Edfors et al. \cite{vanDeBeek1998} obtained a low-rank approximation for the channel autocorrelation matrix by considering only its $ p $ largest singular values, thus reducing the channel estimation complexity. Huang et al. \cite{Willems2007} performed computations in the spatial-domain of multiple-input-multiple-output (MIMO) systems to reduce the complexity of the LMMSE estimator. Noh et al. \cite{Park2006} proposed an approximation to the LMMSE that involved partitioning the autocorrelation matrix into sub-matrices in order to reduce the dimensions of the matrix to be inverted. Ye Li et al. \cite{Sollenberger1998} proposed a channel estimator that is robust to inaccuracies in the knowledge of the channel statistics, while assuming that the time-domain autocorrelation of the channel is characterized by Jakes' model \cite{Jakes1975}.
Despite the advances, LMMSE-based channel estimation algorithms are not a popular choice for practical implementation. Alternative methods for channel estimation focus on transform-domain processing, which involve applying linear or nonlinear functions such as thresholding to the CIR.
%\cite{Huang1997}
Minn and Bhargava \cite{Bhargava2000} proposed an intra-symbol time-averaging based method with most-significant-tap selection to denoise the CIR. Kang et al. \cite{Park2007} used a thresholding function on the CIR to annull taps below a threshold, obtained based on the noise variance. Krondorf et al. \cite{Fettweis2006} proposed a method to estimate the delay-spread of the CIR and used it to annull taps having longer delays than the estimated delay-spread. Yu and Sadeghi {\cite{Sadeghi2012}} proposed a low-complexity method for least-squares (LS) estimation of the CIR using the LS estimate of the frequency-domain pilot subcarriers. Although these algorithms are in general ad hoc, they are less complex than the LMMSE technique and do not require perfect prior knowledge of channel statistics. Their performance is better than the ML technique, but inferior to the LMMSE technique.

%** Contributions of this paper **\\
\subsection{Contributions}
 In this paper, we focus on estimating the CFR from a preamble OFDM symbol. We model the samples of the CFR as a vector of unknown \emph{ deterministic } constants and do not assume prior knowledge of the channel statistics. The proximity of an estimate to the ground truth is quantified by the mean-square error (MSE). Since the ground truth for the estimate of an unknown deterministic constant is not available in practice, the MSE is not computable. Therefore, we replace the oracle MSE with an unbiased estimate known as Stein's unbiased risk estimate (SURE) \cite{Stein1981}, which is dependent only on the noisy observations of the unknown parameters (CFR) and the variance of noise. Next, we select a parametric function to denoise the noisy observations, with the values of the parameters chosen to minimize SURE. 
 
 SURE-based denoising methods have been extensively used in image and speech processing. Luisier et al. \cite{Luisier2007}, Blu and Luisier\cite{Luisier2007Nov}, Raphan and Simoncelli \cite{Raphan2007} proposed methods that use SURE for wavelet denoising in images.  Kishan and Seelamantula \cite{SeelamantulaKishan2012} developed SURE methodology to optimally choose parameters for bilateral filters. Muraka and Seelamantula \cite{Seelamantula2011,Seelamantula2012} derived SURE-optimal frequency-domain based denoising functions and chose parameters for functions that optimize perceptual distortion measures of speech. Zheng et al. \cite{Blu2010} performed denoising of discrete-cosine-transform coefficients using SURE for speech recognition. Krishnan and Seelamantula \cite{Seelamantula2013} used SURE to optimally compute the order of Savitsky-Golay filters for time-varying signals.
 
 Our choice of the parametric denoising function is influenced by the following two properties of the channel: (i) The samples of the CFR are correlated; and (ii) The taps of the CIR with low SNR contribute significantly to the MSE and can be eliminated to reduce the MSE.
 The main contributions of this paper are as follows:
 \begin{itemize}
 \item We have designed a practically implementable algorithm for channel estimation, based on Stein's lemma, that gives a considerable improvement over the ML estimate of the channel. 
 \item The channel estimation algorithm is designed to obtain channel estimates by utilizing the correlation between subcarriers of the CFR and by attenuating the low-SNR taps of the CIR.
 \item Using SURE-based channel estimate for equalization, we show a performance improvement of $ 2.25 $ dB, in SNR, over the ML estimate.
 \end{itemize}

\subsection{Organization of the paper}
In Section \ref{section:systemModel}, we describe the OFDM system model. We briefly review the existing channel estimation algorithms in Section \ref{section:existingChannelEstimationAlgorithms}. In Section \ref{section:proposedMethodology}, we introduce SURE methodology and its application to channel estimation for OFDM systems. We present the results of simulations in Section \ref{section:simulation} and compare the performance of SURE-based channel estimation algorithm with existing algorithms in practical scenarios. Some concluding remarks are given in Section \ref{section:conclusion}.

\subsection{Notations}
We use bold-face lower-case alphabets to denote column vectors $ \mathbf{x} $ and bold-face upper-case alphabets to denote matrices $ \mathbf{X} $. $ {x}_k $ represents the $ k^{th} $ element of the vector $ \mathbf{x} $. $ (\cdot)^* $ , $ (\cdot)^T $ and $ (\cdot)^H $ represent the complex conjugate, transpose, and conjugate transpose (Hermitian) operation, respectively. $ \mathcal{E}\lrc{\cdot} $ denotes the expectation operator and $ \mathcal{E}_{\mathbf{x}}\lrc{\cdot} $ denotes the expectation over the random vector $ \mathbf{x} $. $ \mathbf{I}_N $ represents an $ N \times N $ identity matrix and $ || \cdot || $  denotes the vector norm. $ \mathcal{CN}(\pmb{\mu},\pmb{\Sigma}) $ represents the complex normal distribution with mean $ \pmb{\mu} $ and covariance matrix $ \pmb{\Sigma} $. $ \Re\lrc{\mbf{x}} $ and $ \Im\lrc{\mbf{x}} $ represents the real and imaginary parts, respectively of the complex vector $ \mbf{x} $. $ \text{diag}\lrc{a_1 , \cdots , a_N} $ is used to denote an $ N \times N $ diagonal matrix with $ a_1 , \cdots , a_N $ occupying the diagonal positions.
\section{System model}
\label{section:systemModel}
Consider a cyclic-prefix (CP) OFDM system with $ K $ sub-carriers. If $ \lrc{x_k}_{k=0}^{K-1}$ are the data values transmitted on the $ K $ sub-carriers, the transmitted OFDM symbol can be written as,
\begin{eqnArrAligned*}
	s_n  & = \frac{1}{K} \sum_{k=0}^{K-1} x_k e^{\frac{\mathrm{j}2\pi kn}{K}} \ , \qquad \forall n\in \lrc{-K_g,K-1}, 
\end{eqnArrAligned*}%
where $ \lrc{s_n}_{n=-K_g}^{K-1} $ is the time domain OFDM symbol and $ K_g $ is the length of the CP. If the time-domain OFDM symbol passes through a channel with impulse response $ \mbf{g} $ of length $ P $, the received symbol $ \mbf{r'} $ obtained after discarding the guard interval can be written as,
\begin{dmath}
	\label{eqn:systemModel-timeDomainSystemModel}
	\mbf{r'} = \mbf{g} \otimes \mbf{s} + \pmb{\eta} \ ,
\end{dmath}
where $ \mathbf{s} = [s_0 \ s_1 \cdots \ s_{K-1}]^T $, $ \mathbf{g} = [g_0 \ g_1 \cdots g_{P-1}]^T $ , $ \mathbf{r'} = [r'_0 \ r'_1 \cdots r'_{K-1}]^T $ and $ \otimes $ denotes circular convolution.  $ \pmb{\eta} = [\eta_0 \ \eta_1 \cdots \eta_{K-1}]^T $ is the added white Gaussian noise and $ \pmb{\eta} \sim \mathcal{CN}(\mbf{0},\sigma^2 \mbf{I}_K) $. In \eqref{eqn:systemModel-timeDomainSystemModel}, we have assumed that $ P $, the length of the CIR, is less than $ K_g $, so that there is no loss of orthogonality due to inter-symbol interference. We have also assumed that the synchronization algorithms are accurate and that there is no inter-carrier interference. Multiplying both sides of \eqref{eqn:systemModel-timeDomainSystemModel} by the $ K \times K $ discrete Fourier transform (DFT) matrix $ \mbf{F} $, we get,
\begin{equation}
\label{eqn:systemModel-freqDomainSystemModel}
\mbf{y'} = \mbf{F} \mbf{r'} = \mbf{X} \mbf{h} + \mbf{F} \pmb{\eta} \ , 
\end{equation}
where $ \mathbf{X} = \mathrm{diag}\lrc{x_0,\cdots,x_{K-1}}  $ and $ \mbf{h} $ is the channel frequency response  given by $ \mbf{h} = \mbf{F} \lrs{ \mbf{g}^T \ \\ \overbrace{0 \\ \cdots \\ 0  }^{K-P} }^T$. We restrict ourselves to estimating the channel from a preamble symbol, that is, when the value of $ \mathbf{X} $ is known. We also assume that the symbols $ \lrc{x_0, \cdots ,x_{K-1}} $ and are chosen from a constant modulus constellation set with unit power, so that $ \mathbf{X}^H \mathbf{X} = \mathbf{I}_K $. This assumption is valid in practice, since binary phase-shift keying (BPSK) and quaternary phase-shift keying (QPSK) are preferred modulation schemes for the preamble data, owing to their property of maximum separation between constellation points for a given signal power. Pre-multiplying $ \mbf{y'} $ by $ \mathbf{X}^{-1} $ gives us a noisy observation of the CFR : 
\begin{equation}
\label{eqn:systemModel-channelObservation}
	\mathbf{y} = \mathbf{X}^{-1} \mathbf{y'} = \mathbf{h} +\mathbf{X}^{-1} \mbf{F} \pmb{\eta} = \mathbf{h} + \mbf{w} \ ,
\end{equation}%
where $ \mbf{w} = \mbf{X}^{-1} \mbf{F} \pmb{\eta} $ and $ \mathbf{w} \sim \mathcal{CN}(\mbf{0},\sigma^2(\mathbf{X}^H\mathbf{X})^{-1}) $.
We define the noisy estimate of the CIR as,
\begin{equation} 
\mathbf{r} = \mathbf{F}^H \mathbf{y} = \mbf{F}^H ( \mbf{h} + \mbf{w} ) =  \mbf{g} + \mbf{F}^H \mbf{w} = \mbf{g} + \mbf{v} \ , 
\end{equation}
where $ \mbf{v} = \mbf{F}^H \mbf{w} $. We observe that the channel has two properties, which we utilize to design a channel estimation algorithm :
\begin{enumerate}
\item \emph{The samples of the CFR are correlated.} \\ 
Let $ c_{hh}(k,\ell) \triangleq \mathcal{E}\lrc{h_k h_{k-\ell}^*} $ be the correlation between the channel value at the $ k^{th} $ sub-carrier with that of the $ k-\ell^{th} $ sub-carrier. Assuming that the taps of the CIR are uncorrelated with each other, we find that,
\begin{eqnArrAligned}
\label{eqn:systemModel-channelCorrelation}
	c_{hh}(k,\ell) & = \sum\limits_{p=0}^{P-1} \alpha_p e^{\mathrm{j}\frac{2\pi \ell p}{K}} \qquad \forall \ k \in \lrc{0,1,\cdots,K-1} \ , \\
	& = c_{hh}(\ell),
\end{eqnArrAligned}%
where $ \alpha_p $ corresponds to the average power of the $ p^{th} $ channel tap. From \eqref{eqn:systemModel-channelCorrelation}, we infer that the samples of the channel frequency response are correlated with each other, and that $ c_{hh}(k,\ell) $ is independent of the absolute value of the carrier $ k $ and is only a function of the lag $ \ell $.

\item \emph{The taps of the CIR with a low SNR can be attenuated to improve the MSE.} \\
The MSE between the CIR $ \mbf{g} $ and it's estimate $ \mbf{\widehat{g}} $ is given as,
\begin{dmath}
 	\mbox{MSE} = \frac{1}{K}\mathcal{E}\lrc{||\mbf{g} - \mbf{\widehat{g}}||^2} \ , \\
 	 		   = \frac{1}{K} \sum\limits_{k=0}^{K-1} |g_k - \hat{g}_k|^2  \ , \\
 	 		   = \frac{1}{K} \sum\limits_{k=0}^{K-1} \mbox{MSE}_k \ , \\
\end{dmath}%
where $ \mbox{MSE}_k = \frac{1}{K} \mathcal{E} \lrc{|g_k - \hat{g}_k|^2} $ is the contribution to the MSE from the estimate of the $ k^{th} $ tap of the CIR. The MSE can be reduced by using a thresholding function to attenuate taps with a low SNR, which we demonstrate by means of the following example. If we define the estimate $ \hat{g}_k $ as,
\begin{equation*} \hat{g}_k = a_k r_k = a_k (g_k+v_k), \end{equation*}
where $ a_k \in \lrc{0,1} $, $ \mbox{MSE}_k $ can be written as,
\begin{eqnArrAligned*}
	\mbox{MSE}_k = \begin{cases}
		{|g_k|^2} \ , \qquad  & a _k= 0 \ , \\
		\mathcal{E}\lrc{|v_k|^2} = \sigma^2 \ ,  \qquad & a_k = 1 \ .
	\end{cases}
\end{eqnArrAligned*}
The value of $ a_k \in \lrc{0,1} $ that minimizes MSE can be written as,
\begin{eqnArrAligned}
\label{eqn:systemModel-lowSnrAttenuationProperty}
	a_k^{\text{opt}} = \begin{cases}
		1 \ , \qquad & \frac{{|g_k|^2}}{\sigma^2} > 1 \ , \\ 
		0 \ , & \mbox{otherwise} \ .
	\end{cases}
\end{eqnArrAligned}%
From \eqref{eqn:systemModel-lowSnrAttenuationProperty}, we infer that a reduction in MSE can be obtained using a thresholding function and the magnitude of reduction is $ \sigma^2 - {|g_k|^2} $ when the SNR of the $ k^{th} $ tap (given by $ \frac{{|g_k|^2}}{\sigma^2} $) is less than $ 1 $.
\end{enumerate}

\subsection{Brief review of existing channel estimation algorithms}
\label{section:existingChannelEstimationAlgorithms}
\subsubsection{Maximum-likelihood channel estimation}
\label{SetA1/p1/a7_channelEstimationAlgorithms/subsection:mle}
The maximum-likelihood estimation of the channel, when the noise is white and Gaussian distributed, reduces to
\begin{eqnArrAligned}
\label{eqn:mle-optimEqn}
\widehat{\mathbf{h}}_{\text{mle}} & = \arg \min_{\mathbf{h}} \ || \mathbf{y} - \mathbf{h} ||^2 \\
\widehat{\mathbf{h}}_{\text{mle}} & = \mathbf{y} \ ,
\end{eqnArrAligned}%
that is, that the  maximum-likelihood estimate of the channel is the noisy observation itself \cite{Arslan2007}.

\subsubsection{Linear minimum mean-square error channel estimation}
\label{SetA1/p1/a7_channelEstimationAlgorithms/subsection:lmmse}
The LMMSE approach for channel estimation requires complete knowledge of the channel statistics to obtain a set of linear coefficients that minimize the MSE. The LMMSE estimate of the channel is given as \cite{AliSayed2008}
\begin{eqnArrAligned*}
\label{eqn:lmmse}
\widehat{\mathbf{h}}_{\text{lmmse}} & = \mathbf{A}_{\text{lmmse}} \ \mathbf{y} \ , \\
\end{eqnArrAligned*}%
with
\begin{eqnArrAligned}
					\mathbf{A}_{\text{lmmse}} & = (\mathbf{C}_{hh} + \sigma^2 \mathbf{I}_K)^{-1} \mathbf{C}_{hh} \ ,\\ 
\end{eqnArrAligned}%
where $ \mathbf{C}_{hh} \triangleq \mathcal{E}_{h}\lrc{\mathbf{h} \ \mathbf{h}^H} $ is the autocorrelation matrix of the channel frequency response.
\subsubsection{Channel estimation using CIR thresholding}
In \cite{Park2007} Kang et al. proposed a thresholding-based approach for denoising the observations of the channel and calculated the threshold as,
\begin{equation}
	T = 2 \hat{\sigma}^2 \ ,
\end{equation}
where $ \hat{\sigma}^2 $ is an estimate of the noise variance computed from the observations. The thresholding function is defined as
\begin{eqnArrAligned}
\hat{g}_k = \begin{cases}
	r_k \ , \quad ||r_k||^2 \geq T \ , \\
 	0 \ , \quad   ||r_k||^2 < T \ ,
\end{cases}
\end{eqnArrAligned}
and $ \hat{g}_k $ is an estimate of the $ k^{th} $ tap of the CIR.
\section{Proposed methodology}
\label{section:proposedMethodology}
We devise a parametric function to denoise the measurements and obtain the channel estimate. Rather than optimizing the parameters of the denoising function with respect to the mean-square estimation error, which is not feasible in the classical parameter estimation regime, we optimize an unbiased estimate of the mean-square error, in particular, SURE. We consider both linear and nonlinear forms of the denoising function and compare their relative performance.

\subsection{Mathematical preliminaries}

For a $ K \times 1 $ vector $ \mathbf{x} $ and its estimate $ \mathbf{\hat{x}} $, the mean-square error is defined as,
\begin{eqnArrAligned*}
	\mbox{MSE} & = \mathcal{E}\lrc{\frac{1}{K}||\mathbf{x} - \mbf{\hat{x}}||^2} \\ 
		& = \lrc{\frac{1}{K}\sum_{k=0}^{K-1} \mathcal{E}| x_k - \hat{x}_k|^2  } \ .
\end{eqnArrAligned*}%
For the case where we observe the unknown parameter $ \mathbf{x} $ in white Gaussian noise, that is $ \mathbf{y} = \mathbf{x} + \mathbf{w} $, with $ \mathbf{w} \sim \mathcal{N}(\mbf{0},\sigma^2\mathbf{I}_K) $, the multidimensional version of Stein's lemma \cite{Stein1981} allows for an unbiased estimate of the MSE from the observations $ \mathbf{y} $. The scalar version of Stein's lemma, which can be extended to the multidimensional case, is reproduced next.
\begin{lemma}[Stein, 1981 \cite{Stein1981}]
\label{lemma:steinsLemma}
Let $ w $ be a $ \mathcal{N}(0,1) $ real random variable and let $ f : \mathbb{R} \rightarrow \mathbb{R} $ be an indefinite integral of the Lebesgue measurable function $ f' $, essentially the derivative of $ f $. Suppose also that $ \mathcal{E}\lrs{|f'(w)| < \infty} $. Then
\begin{eqnArrAligned*}
	\mathcal{E}(f'(w)) = \mathcal{E}(wf(w))
\end{eqnArrAligned*}
\end{lemma}%
Since $ \mbf{\hat{x}} $ is obtained from the observation vector $ \mbf{y} $ using a denoising function $ f(\cdot) $, we write $ \mbf{\hat{x}} = f(\mbf{y})$. Lemma \ref{lemma:steinsLemma} can be extended for the case when  $ f:\mathbb{C}^K \rightarrow\mathbb{C}^K $ as,
\begin{equation}
\label{eqn:sureMultidimensionalVersion}
	\mathcal{E}\lrc{\mbf{w}^H \mbf{\hat{x}}} = 	\mathcal{E}\lrc{\mbf{w}^H f(\mbf{y})} = \mathcal{E}\lrc{\nabla . \ f(\mbf{y})} \ ,
\end{equation}%
where the divergence term is given by
\begin{equation}
	\label{eqn:sureDivergenceTerm}
	\nabla . f(\mbf{y}) = \sum\limits_{k=0}^{K-1} \frac{\partial f_k(\mbf{y})}{\partial y_k} \ ,
\end{equation}%
with
\begin{equation}
\hat{x}_k = f_k(\mbf{y}) \ .
\end{equation}%
Using \eqref{eqn:sureMultidimensionalVersion},  we define a random variable $ {\epsilon} $ \cite{Luisier2007,Luisier2007Nov} as,
\begin{dmath}
\label{eqn:sureVectorCostFunction}
\epsilon \triangleq \frac{1}{K}\lrs{||\mathbf{y}||^2 - K\sigma^2 + ||\mathbf{\hat{x}}||^2 - 2\Re\lrc{\mathbf{y}^H\mathbf{\hat{x}}} + 2\sigma^2\Re\lrc{\nabla. \ \hat{\mathbf{x}}}} \ , \\ 
= \frac{1}{K}\lrs{||\mathbf{y}||^2 - K\sigma^2 + ||f(\mathbf{y})||^2 - 2\Re\lrc{\mathbf{y}^H f(\mathbf{y})} + 2\sigma^2\Re\lrc{\nabla. \ f({\mathbf{y}})}} \ .
 \end{dmath}%
$ {\epsilon} $ is Stein's unbiased risk estimator and has a mean value equal to the MSE, as shown in (\ref{eqn:averageValueOfSure}).
\begin{dmath}
\label{eqn:averageValueOfSure}
\mathcal{E}\lrc{\epsilon} = \mathcal{E} \lrc{\frac{1}{K}\lrs{||\mathbf{y}||^2 - K\sigma^2 + ||f(\mathbf{y})||^2 - 2\Re\lrc{\mathbf{y}^H f(\mbf{y})} + 2\sigma^2\Re\lrc{\nabla. \ f(\mathbf{y})}}}
= \frac{1}{K} \lrc{||\mathbf{x}||^2 + \mathcal{E}\lrc{||f(\mbf{y})||^2} -2 \Re\lrc{\mathcal{E}\lrc{\mathbf{y}^Hf(\mathbf{y})} } + 2\Re\lrc{\mathcal{E}\lrc{(\mathbf{y} - \mathbf{x})^Hf({\mathbf{y}})}}} \\
= \frac{1}{K} \lrc{||\mathbf{x}||^2 + \mathcal{E}\lrc{||\mbf{\hat{x}}||^2} -2 \Re\lrc{\mathcal{E}\lrc{\mathbf{y}^H\mbf{\hat{x}}} } + 2\Re\lrc{\mathcal{E}\lrc{(\mathbf{y} - \mathbf{x})^H\mbf{\hat{x}}}}} \\
= \mathcal{E}\lrc{\frac{1}{K}||\mathbf{x} - \hat{\mathbf{x}}||^2}
= \mbox{MSE} \ .
\end{dmath}%
$ {\epsilon} $ is dependent only on the observations $ \mathbf{y} $ and its variance is inversely proportional to the square of the number of observations $ K $ \cite{Pesquet2009}. For large values of $ K $, SURE can be treated as equivalent to the oracle MSE in obtaining the optimal parameters of the denoising function.
\subsection{SURE-optimized linear estimate of the channel}
\label{subsection : SURE Optimized Linear Estimate of the Channel}
We use the correlation between sub-carriers in (\ref{eqn:systemModel-channelCorrelation}) to define the linear denoising function while treating the channel coefficients $ \mathbf{h} $ as a vector of unknown deterministic constants. We take the channel estimate of the $ k^{th} $ sub-carrier to be a weighted linear combination of the observations on $ L $ sub-carriers on either side of the $ k^{th} $ sub-carrier:
\begin{eqnArrAligned}
\label{eqn:sure-estimationEqn}
\hat{h}_k & =  \sum_{\ell=-L}^{L} a_\ell . r_{(k+\ell)_K} \qquad \forall k \in \{0,1,\cdots,K-1\} \ , \\
\end{eqnArrAligned}%
where $ (\cdot)_K $ denotes the modulo-$ K $ operation. This procedure is repeated for all $ k \in \{0,1,\cdots,K-1\} $ to obtain the estimate for each of the $ K $ sub-carriers. Stacking these estimates into a vector, we get the channel estimate $ \hat{\mathbf{h}} $.
Further, the fact that $ c_{hh}(k,\ell) = c_{hh}(\ell) $, allows us to use a single set of weighting coefficients $ \mathbf{a} $ for every sub-carrier. Therefore, we redefine the channel estimate using matrix notation as,
\begin{eqnArrAligned}
\label{eqn:SURE-generalLinearEstimationEqn}
\mbf{\widehat{h}} & =  \mathbf{Y} \mathbf{a} \ , \\
\end{eqnArrAligned}%
where
\begin{eqnArrAligned}
\label{eqn:sure-circShiftedChannelObservations}
\mathbf{Y}_{K \times N} & = \begin{bmatrix}
		y_{(-L)_K} 	 	 & y_{(-L+1)_K} & \ldots  	& y_{(0)_K}   	& \ldots & 	y_{(L-1)}	& 	y_{(L)_K} \\
		y_{(-L+1)_K} 	 & y_{(-L+2)_K} & \ldots  	& y_{(1)_K}   	& \ldots & 	y_{(L)}	 	&	y_{(L+1)_K} \\
		\vdots		 	 & \vdots 		& \ddots 	& \vdots     	& \ddots &  \vdots	 	& 	\vdots \\		
		y_{(-L+(K-1))_K} & y_{(-L+K)_K}	& \ldots 	& y_{(K-1)_K} 	& \ldots &	y_{(L+K-2)}	& 	y_{(L+(K-1))_K} \\				
		 \end{bmatrix}
\end{eqnArrAligned}
and
\begin{equation*}
		\mathbf{a} \triangleq \left[a_{-L}, \cdots ,a_0 , \cdots ,a_L\right]^T \ .
\end{equation*}%
If $ \mbf{P} $ is the cyclic-permutation matrix,
\begin{equation*}
\mbf{P}_{K\times K} = \left[
\begin{array}{c c c c}
  0 & \cdots & 0 & 1 \\
& \raisebox{-12pt}{{\Large\mbox{{$\mbf{I}_{K-1}$}}}} & & 0 \\[-4ex]
&&&\vdots\\[-0.5ex]
&&& 1
\end{array}
\right] \ ,
\end{equation*}
\eqref{eqn:sure-circShiftedChannelObservations} can be rewritten as
\begin{equation*}
		 \mbf{Y} = \begin{bmatrix}
		\mbf{P}^{L} \mbf{y} & \vdots & \mbf{P}^{L - 1} \mbf{y} & \vdots & \cdots & \mbf{P}^{0} \mbf{y} & \cdots & \vdots& \mbf{P}^{-L+1} \mbf{y} & \vdots & \mbf{P}^{-L} \mbf{y}
		 \end{bmatrix}\\ \ .
\end{equation*}
By parameterizing the channel estimation function, we have reduced the problem of estimating the channel values in $ K $ sub-carriers to estimating $ N $ weighting coefficients, $ \mathbf{a} $ that minimize SURE.
\begin{dmath}
\label{eqn:SURE-optimEqn}
\mathbf{a}_{\text{sure}}  =  \arg \min_{\mathbf{a}} \lrc{\epsilon}  \\ 
= \arg \min_{\mathbf{a}} \frac{1}{K} \lrc{||\mathbf{y}||^2 - K\sigma^2 + \mathbf{a}^H \mathbf{Y}^H \mathbf{Y} \mathbf{a} - 2 \Re \lrc{\mbf{y}^H \mbf{Y} \mbf{a}} + 2 \sigma^2 \Re \lrc{\nabla \ . \mbf{Y} \mbf{a}} } \ , \\
= \arg \min_{\mathbf{a}} \frac{1}{K} \lrc{||\mathbf{y}||^2 - K\sigma^2 + \mathbf{a}^H \mathbf{Y}^H \mathbf{Y} \mathbf{a} - 2 \Re \lrc{\mbf{y}^H \mbf{Y} \mbf{a}} + 2 \sigma^2 \Re \lrc{\mbf{b}^H \mbf{a}}}  \ , \\
\end{dmath}
where
\begin{equation*}
 \mathbf{b}_{N\times 1} \triangleq \lrs{\overbrace{0,\cdots,0}^{\text{L}},M,\overbrace{0,\cdots,0}^{\text{L}}}^T  \\ 
\end{equation*}%
and
\begin{equation*}
N \triangleq 2L+1 \ .
\end{equation*}
%Note that $ \left\{\mathbf{a}^H R^H R \mathbf{a} \right\} $ and $ \left\{ R^H \mathbf{r} - \sigma^2 \mathbf{b} \right\} $ are the unbiased estimates of $ \mathbb{E} \{\mathbf{a}^H R^H R \mathbf{a}\} $ and $ \mathbb{E} \{R^H \mathbf{h}\} $, respectively. 
The expression for the optimal weights turns out to be a solution to the following system of equations: 
\begin{eqnArrAligned}
\label{eqn:SURE-optimWeights}
{\mathbf{Y}^H\mathbf{Y}} \ \mathbf{a}_{\text{sure}} & =   \mathbf{Y}^H \mathbf{y} - \sigma^2 \mathbf{b} \\
\mathbf{a}_{\text{sure}} & = \left(\mathbf{Y}^H \mathbf{Y} \right)^{-1} \left[ \mathbf{Y}^H \mathbf{y} - \sigma^2 \mathbf{b}\right] \ . \\
\end{eqnArrAligned}%
For a circulant matrix with elements of the first column Gaussian distributed and i.i.d., the eigenvalues are also Gaussian distributed and i.i.d. The probability of this random matrix being singular is equal to the probability that at least one of the eigenvalues is $ 0 $, which is $ 0 $. Since the columns of $ \mbf{Y} $ are a subset of the columns of a circulant matrix, they are linearly independent and therefore, $ \mathbf{Y}^H \mathbf{Y} $ is of full rank with probability $ 1 $.
The resulting estimate of the channel using SURE optimized coefficients is given by
\begin{eqnArrAligned}
\label{eqn:SURE-linearEstimationEqn}
	\mbf{\widehat{h}}_{\text{sure}} = \mathbf{Y} \ \mathbf{a}_{\text{sure}} \ .
\end{eqnArrAligned}%
We also observe that a special case of this formulation, with $ N = 1 $, gives rise to the James-Stein estimator \cite{Stein1961} :
\begin{eqnArrAligned*}
{a}_{\text{sure}}^{\text{JS}} = 1 - \frac{K\sigma^2}{||\mathbf{y}||^2} \ .
\end{eqnArrAligned*}
\subsection{SURE-optimized nonlinear estimate of the channel}
\label{subsection:SURE Optimized Non-linear Estimate of the Channel}
To improve the MSE performance, we augment the linear denoising function in \eqref{eqn:SURE-generalLinearEstimationEqn} with a thresholding function to attenuate taps of the CIR that are associated with a low SNR. 

The estimate of the channel is rewritten as,
\begin{eqnArrAligned}
\label{eqn:nonLinearSureDefinition}
\mbf{\widehat{h}} &= \mathbf{Y}\mathbf{a} + a_{L+1}\mathbf{y_T} \ , \\
\end{eqnArrAligned}
where
\begin{eqnArrAligned}
\label{eqn:nonLinearThresholdingFunction}
\mathbf{y_T} &= \mathbf{F} \ \mathbf{r_T} \ , \\
\mbf{r_T} &= \mbf{r} - q(\mbf{r}) \ , \\ 
q(\mathbf{r}) & = \begin{bmatrix} q_1(r_1) & q_2(r_2) & \cdots & q_{K-1}(r_{K-1}) \end{bmatrix}^T \ ,
\end{eqnArrAligned}
and $ \lrc{q_k(r_k)}_{k=0}^{K-1}$ is a collection of scalar point-wise thresholding functions operating on the noisy observations of the CIR. Equation \eqref{eqn:nonLinearThresholdingFunction} is equivalent to the convex combination of the linear denoising function \eqref{eqn:SURE-generalLinearEstimationEqn} and the DFT of the thresholding function $ q(\mbf{r}) $.

\begin{figure}[t!]
\centering
	\begin{tikzpicture}
		% The following line determines the location of the legend
% E.g. - 1 : \pgfplotsset{every axis legend/.append style={legend pos = south west} , legend cell align = {left}}
% In E.g.2, the legend is placed at point (x,y) and the corner of the legend that sits on (x,y) is chosen by the anchor
% E.g. - 1 : \pgfplotsset{every axis legend/.append style={at={(0,0)},anchor=south west} , legend cell align = {left}}

\pgfplotsset{every axis legend/.append style={legend pos = north west},legend cell align = {left}}
        
% The following setting allows font size choices. Options are \tiny, \footnotesize, \small, \normalsize, \large, \Large, \LARGE, \huge, \Huge
\pgfplotsset{%
tick label style={font=\normalsize},
title style={font=\normalsize,align=center},
label style={font=\normalsize,align=center},
legend style={font=\footnotesize},
xticklabels={,,}
            }

\begin{axis}[%
xlabel = {$r_k$},
ylabel = {$q_k(r_k)$},
%title  = {Plot of various point-wise thresholding functions \\ for a real input},
enlargelimits = true,
cycle list name = {color},
grid = major,
scale = 1]

\addlegendentry{$q_k(r_k) = r_k$}

\addplot [ color = {blue}, 
           mark  = {none},
           style = {dashed},
           line width = 1pt,
           smooth] table {\figDir/datFiles/fig25-line1.dat};

\addlegendentry{$q^S_k(r_k)$}

\addplot [ color = {red}, 
           mark  = {none},
           style = {dashdotted},
           line width = 1pt] table {\figDir/datFiles/fig25-line2.dat};

\addlegendentry{$q^H_k(r_k)$}

\addplot [ color = {green}, 
           mark  = {none},
           style = {densely dotted},
           line width = 1pt] table {\figDir/datFiles/fig25-line3.dat};

\addlegendentry{$q^{\text{LET}}_k(r_k)$}

\addplot [ color = {black}, 
           mark  = {none},
           style = {solid},
           line width = 1pt] table {\figDir/datFiles/fig25-line4.dat};

\end{axis}

\draw (4.25,2.40) node[anchor = south west,align=left]{T};
\draw (1.95,2.75) node[anchor = south west,align=left]{-T};
	\end{tikzpicture}
	\caption{Plot of various point-wise thresholding functions for a real parameter input.}
	\label{fig:pointwiseThresholdingFunctions}
\end{figure}
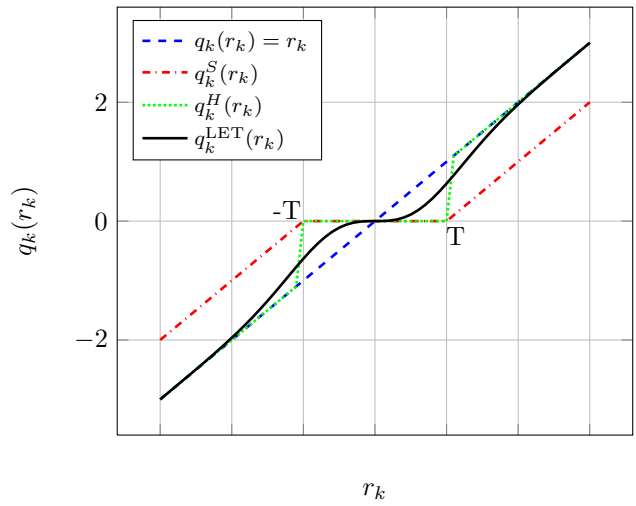

We experimented with the following nonlinear point-wise thresholding functions:
\begin{itemize}
\item \textbf{Hard-thresholding} : $ q^H_k({r}_k) = \begin{cases}
{r}_k \ , & |{r}_k| \geq T  \ ,\\
0 \ , & |{r}_k| < T \ , \\
\end{cases} $
\item \textbf{Soft-thresholding} : $ q^S_k({r}_k) = \max\lrc{|{r}_k| - T,0} e^{j\angle{{r}_k}} $ \ ,
\item \textbf{Linear Expansion of Thresholds (LET) \cite{Luisier2007} : } $ q^{\text{LET}}_k({r}_k) = {r}_k \left( 1 - e^{-\frac{|{r}_k|^2}{T}} \right) \ , $ 
\end{itemize}
where $ T $ is the threshold.
In Fig. \ref{fig:pointwiseThresholdingFunctions}, we plot the hard thresholding, soft thresholding and LET functions for a real parameter input. 

\begin{figure}[t!]
\centering
	\begin{tikzpicture}
		% The following line determines the location of the legend
% E.g. - 1 : \pgfplotsset{every axis legend/.append style={legend pos = south west} , legend cell align = {left}}
% In E.g.2, the legend is placed at point (x,y) and the corner of the legend that sits on (x,y) is chosen by the anchor
% E.g. - 1 : \pgfplotsset{every axis legend/.append style={at={(0,0)},anchor=south west} , legend cell align = {left}}

\pgfplotsset{every axis legend/.append style={legend pos = north east},legend cell align = {left}}
        
% The following setting allows font size choices. Options are \tiny, \footnotesize, \small, \normalsize, \large, \Large, \LARGE, \huge, \Huge
\pgfplotsset{%
tick label style={font=\normalsize},
title style={font=\normalsize,align=center},
label style={font=\normalsize,align=center},
legend style={font=\footnotesize}
            }

\begin{axis}[%
xlabel = {T},
ylabel = {MSE of the channel estimate},
%title  = {Plot of MSE for different values of $T$ and $p$ },
enlargelimits = true,
cycle list name = {color},
grid = major,
smooth,
scale = 1]

\addlegendentry{p = 2}

\addplot [ color = {blue}, 
           mark  = {square},
           style = {solid},
           line width = 1pt] table {\figDir/datFiles/fig27-line1.dat};

\end{axis}
%\draw(1.000000e-01,4.000000e-01) node[anchor = south west,align=left]{ Number of OFDM carriers - 2048\\ 
%$SNR = 10dB$};
;		
	\end{tikzpicture}
	\caption{MSE of the 3GPP channel estimate for various values of $ T $ with $ K = 2048 $, $ p = 2 $, $ \sigma^2 = 0.1 $}
	\label{fig:mseForNonLinearSureParamVariation}
\end{figure}
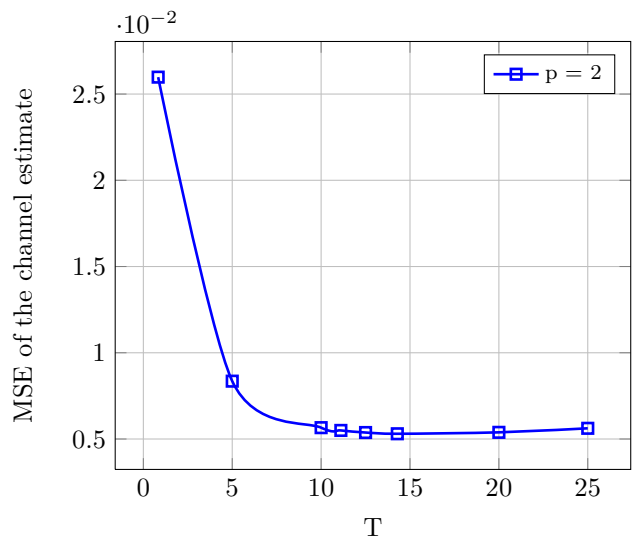

Since Stein's lemma is defined only for weakly differentiable functions that have a bounded-derivative, the usage of a hard-thresholding function is ruled out. Alternatively, soft-thresholding has bounded weak derivatives but does not lend itself to a closed-form solution and requires nonlinear optimization with respect to $ T $. To work around these hurdles, we resort to the LET function. In \eqref{eqn:nonLinearSureDefinition}, we augment \eqref{eqn:SURE-generalLinearEstimationEqn} with the LET function using a weight $ a_{L+1} $. We choose $ a_{L+1} $ in conjunction with weights $ a_{-L},\ldots,a_L $ to annul taps that are associated with a low SNR.
Define
\begin{eqnArrAligned*}
\mathbf{a}^\dagger&  \triangleq  \lrs{\overbrace{a_{-L}, \ldots, a_L}^{\mathbf{\text{a}}} \quad a_{L+1} }_{(N+1)\times 1} \ , \\ 
\end{eqnArrAligned*}
and
\begin{eqnArrAligned*}
\mathbf{Y_T} & \triangleq [\mathbf{Y} \ \vdots \ \mathbf{y_T}]_{M\times(N+1)} \ ,
\end{eqnArrAligned*}%
where $ \mbf{y_T} $ is obtained from \eqref{eqn:nonLinearThresholdingFunction} with $ \mbf{r_T} = q^{\text{LET}}(\mbf{r}) $. We obtain the optimum values of $ \mbf{a^\dagger} $ and threshold $ T $ by minimizing SURE. To simplify the optimization, we fix $ T $ and minimize over the remaining $ N+1 $ parameters. The SURE cost function from (\ref{eqn:sureVectorCostFunction}) for a given value of $ T $ is written as,
\begin{eqnArrAligned*}
\epsilon^{(T)} &= \frac{1}{K} ||\mathbf{\widehat{h}} - \mathbf{h} ||^2 \\
			 &= \frac{1}{K} \lrc{\lrc{\mathbf{Y_T} \mathbf{a}^\dagger}^H \mathbf{Y_T} \mathbf{a}^\dagger - 2\Re\lrc{\mathbf{r}^H \mathbf{Y_T} \mathbf{a}^\dagger} + 2\sigma^2 \Re\lrc{\nabla . \ \mbf{\widehat{h}}}} \ . \\
\end{eqnArrAligned*}%
We consider the values of $ T $ in the range $ (0,25\sigma^2) $. The i.i.d. property of the added noise ensures that $ \left({{\mathbf{Y_T}}^H \mathbf{Y_T}}\right) $ is invertible with probability 1. $ \epsilon^{(T)} $ is simplified as,
\begin{dmath}
\label{eqn:sureNonLinearCostFunction}
\epsilon^{(T)} 	=  \frac{1}{K} \left|\left| \left({{\mbf{Y_T}}^H \mbf{Y_T}}\right)^{\frac{1}{2}}\mathbf{a}^\dagger - \left({{\mbf{Y_T}}^H \mbf{Y_T}}\right) ^{-\frac{1}{2}} \lrc{{\mbf{Y_T}}^H\mathbf{r} - \sigma^2 \mbf{\pmb{\beta}} } \right|\right|^2  - \frac{1}{K} \left({\mbf{Y_T}}^H\mathbf{r} - \sigma^2\mbf{\pmb{\beta}}\right)^H\left({{\mbf{Y_T}}^H \mbf{Y_T}}\right) ^{-1} \left({\mbf{Y_T}}^H\mathbf{r} - \sigma^2 \mbf{\pmb{\beta}} \right) \ , \\
\end{dmath}
where
\begin{eqnArrAligned*}
 \nabla.\mbf{\widehat{h}} &= \pmb{\beta}^H \mbf{a^\dagger} \ , \\
 \mbf{\pmb{\beta}}_{(N+1)\times1} &\triangleq \lrs{\overbrace{0,\ldots,0}^{\text{L}},M,\overbrace{0,\ldots,0}^{\text{L}} \quad \nabla_r.\mathbf{r_T}}^T \ ,
\end{eqnArrAligned*}
and
\begin{eqnArrAligned*}
 \nabla_r.\mathbf{r_T}& = \sum_{k=0}^{K-1} \frac{\partial \lrc{q^{\mathrm{LET}}_k(r_k)}}{\partial r_k} = \sum_{k=0}^{K-1} \lrs{\exp\lrc{-\frac{|r_k|^2}{T}}\left(1 - \frac{|r_k|^2}{T}\right)} \ .
\end{eqnArrAligned*}%
From \eqref{eqn:sureNonLinearCostFunction}, we see that $ \epsilon^{(T)} $ attains a minimum when
\begin{eqnArrAligned}
\mathbf{a^\dagger}_{\text{sure}}^{(T)} & =  \left({\mbf{Y_T}}^H\mbf{Y_T}\right)^{-1} \lrs{\mbf{Y_T}^H\mathbf{r} - \sigma^2 \mbf{\pmb{\beta}}} \ . \\
\end{eqnArrAligned}%
However, the computation of the optimal value of $ T $ is a non-convex problem. In Fig. \ref{fig:mseForNonLinearSureParamVariation}, we show the variation of the MSE of the channel estimate with $ T $ in 3GPP (Third Generation Partnership Project) typical urban channel scenario. A similar behaviour was observed for other channel scenarios that we considered. Based on these observations, we hypothesize that empirically setting $ T $ to a constant (between $ 10 \sigma^2 $ and $ 15 \sigma^2 $) is sufficient for all scenarios of the channel, OFDM symbol sizes, and SNR. Sophisticated optimization with respect to $ T $ was not found to yield commensurate gains.

%Note that the dependence of $ \lrc{H^\dagger} $ over $ T $ is hidden in the above equation. The $ N+2 $ dimensional search is now reduced to a one dimensional search over $ T $. Now, to find the optimal value of $ T $, we need to minimize (\ref{eqn:sureNonLinearCostFunction}) over $ T $. 
%\begin{eqnArrAligned}
%\label{eqn:sureOptimalT}
%T_{sure} 	&= \min_T \quad \epsilon^{(T)} \\ 
%			&= \min_T \quad - \left(\lrc{H^\dagger}^H\mathbf{r} - \sigma^2\mbf{\pmb{\beta}}\right)^H\left({\lrc{H^\dagger}^H H^\dagger}\right) ^{-1} \\
%			& \quad .\left(\lrc{H^\dagger}^H\mathbf{r} - \sigma^2 \mbf{\pmb{\beta}} \right) \\
%			&= \min_T \quad S^{-1}\left|v - \mathbf{v}^H \left(H^H H\right)^{-1} H^H \mathbf{h}^\dagger \right|^2 \\
%\text{where, } \\
%	\mathbf{u} & = \lrs{\lrc{H^\dagger}^H\mathbf{r} - \sigma^2 e_{L+1}} \\
%	v 		 & = \mathbf{h}^\dagger \mathbf{r} - \sigma^2 \ \nabla.\mathbf{g}^\dagger
%\end{eqnArrAligned}
%And $ S $ is the schur complement of the matrix $ \lrc{H^\dagger}^HH  $
%defined as $ 	S = \left|\left|\mathbf{h}^\dagger\right|\right|^2 - \lrc{\mathbf{h}^\dagger}^H H \left(H^H H\right)^{-1} H^H \mathbf{h}^\dagger $. With these, the optimum value of $ T $ is obtained as the minimizer of the scalar function in (\ref{eqn:sureOptimalT}). However, the minimizing function is non-convex. In addition, the variance $ T_{sure} $ is high requiring a large value of $ K $.
\section{Simulation Results}
\label{section:simulation}

We compare the mean-square error performance of the channel estimates based on the LMMSE, CIR-thresholding \cite{Park2007} maximum-likelihood criterion, and SURE criterion. We also compare the bit-error-rate (BER) performance of the channel estimation methods, post equalization, for various channel scenarios, namely, the AWGN environment, single-tap Rayleigh fading channel and the 3GPP typical urban channel scenario \cite{3GPPChannelModel}.

\subsection{Simulation setup}
We simulate the considered channel estimation algorithms for OFDM symbol sizes $ K = 64,256, $ and $ 1024 $. 
%For SURE-based channel estimate we have simulated for $ N = 3, 5 $. 
To render SURE-based channel estimate completely data-dependent, the true value of the noise variance $ \sigma^2 $ has been replaced by an estimate $ \hat{\sigma}^2 $ that is obtained from 500 blank OFDM carriers. In computing the BER and the MSE, $ 20,000 $ trials were simulated for each SNR value. For the BER simulation, data was taken from a 16 quadrature amplitude modulation (QAM) constellation and encoded with a rate 1/2 convolutional code with generator polynomial $ G_1(D) = 1 + D^3 + D^4 + D^5 + D^6 $ and $ G_2(D) = 1 + D^3 + D^4 + D^6  $. The receiver used a symbol-by-symbol maximum a posteriori probability (MAP) estimator with hard-decision Viterbi decoding.

From our simulation, we observed that setting $ T = 12\sigma^2 $ for SURE-based nonlinear channel estimation algorithm provided optimal results for various channel scenarios.

\subsection{Simulation results}
\begin{figure*}[t]
\centering
	{
		\subfigure[] %[Comparison of the mean square error performance for the single-tap Rayleigh channel]
			{	
				\begin{tikzpicture}
					% The following line determines the location of the legend
% E.g. - 1 : \pgfplotsset{every axis legend/.append style={legend pos = south west} , legend cell align = {left}}
% In E.g.2, the legend is placed at point (x,y) and the corner of the legend that sits on (x,y) is chosen by the anchor
% E.g. - 1 : \pgfplotsset{every axis legend/.append style={at={(0,0)},anchor=south west} , legend cell align = {left}}

\pgfplotsset{every axis legend/.append style={legend pos = north east},legend cell align = {left}}
        
% The following setting allows font size choices. Options are \tiny, \footnotesize, \small, \normalsize, \large, \Large, \LARGE, \huge, \Huge
\pgfplotsset{%
tick label style={font=\normalsize},
title style={font=\normalsize,align=center},
label style={font=\normalsize,align=center},
legend style={font=\normalsize}
            }

\begin{axis}[%
xlabel = {SNR (dB)},
ylabel = {MSE of the channel estimate},
%title  = {MSE Comparison for the single-tap Rayleigh channel scenario},
enlargelimits = true,
cycle list name = {color},
grid = major,
smooth,
scale = 1]

\addlegendentry{MLE}

\addplot [ color = {blue}, 
           mark  = {asterisk},
           style = {solid},
           line width = 1pt] table {\figDir/datFiles/fig3-line1.dat};

\addlegendentry{LMMSE}

\addplot [ color = {cyan}, 
           mark  = {square},
           style = {solid},
           line width = 1pt] table {\figDir/datFiles/fig3-line2.dat};

\addlegendentry{SURE Linear : $N = 3$}

\addplot [ color = {black}, 
           mark  = {o},
           style = {solid},
           line width = 1pt] table {\figDir/datFiles/fig3-line3.dat};

\addlegendentry{SURE Nonlinear : $N = 3$}

\addplot [ color = {magenta}, 
           mark  = {diamond},
           style = {solid},
           line width = 1pt] table {\figDir/datFiles/fig3-line4.dat};

\addlegendentry{CIR Thresholding \cite{Park2007}}

\addplot [ color = {red}, 
           mark  = {+},
           style = {solid},
           line width = 1pt] table {\figDir/datFiles/fig3-line5.dat};

\end{axis}
				\end{tikzpicture}
				\label{fig:mseRayleigh}
			}	
		\hspace{0.5cm}
		
		\subfigure[] %[Comparison of the bit error rate performance for the single-tap Rayleigh channel scenario]
		{	
			\begin{tikzpicture}
				\input{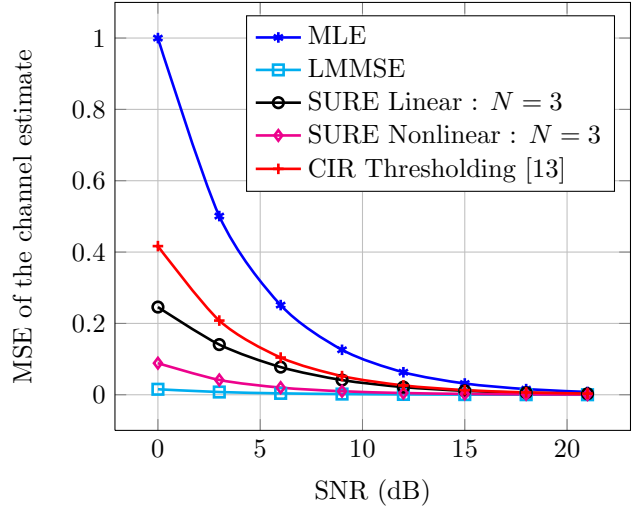}
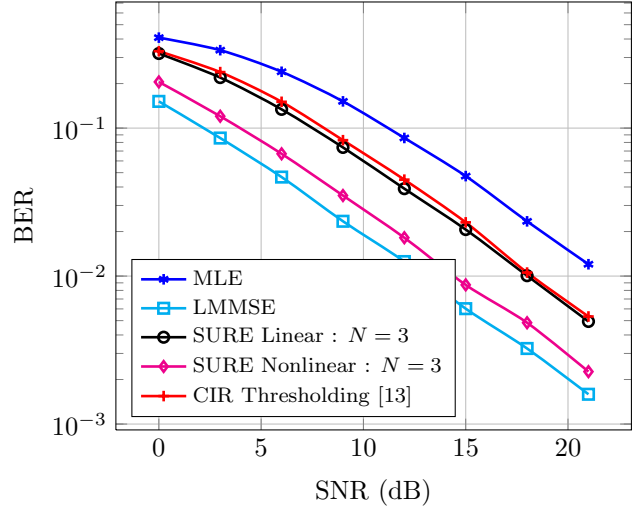
			\end{tikzpicture}
				\label{fig:berRayleigh}			
		}		
	}
\caption{ Comparison of \subref{fig:mseRayleigh} mean-square error performance and \subref{fig:berRayleigh} bit error rate performance between the LMMSE estimate, the ML estimate, CIR-thresholding \cite{Park2007} and the SURE-based (linear and nonlinear) estimate with $ N = 3 $, OFDM symbol size $ K = 64 $ and 16-QAM constellation,. The simulation has been performed for the single-tap Rayleigh fading channel scenario.}
\label{fig:mseCurvesVariousChannels}
\end{figure*}

\begin{figure*}[t]
\centering
	{	

		\subfigure[] %[Comparison of the mean square error performance for 3GPP typical urban channel]
		{
			\begin{tikzpicture}
				% The following line determines the location of the legend
% E.g. - 1 : \pgfplotsset{every axis legend/.append style={legend pos = south west} , legend cell align = {left}}
% In E.g.2, the legend is placed at point (x,y) and the corner of the legend that sits on (x,y) is chosen by the anchor
% E.g. - 1 : \pgfplotsset{every axis legend/.append style={at={(0,0)},anchor=south west} , legend cell align = {left}}

\pgfplotsset{every axis legend/.append style={legend pos = north east},legend cell align = {left}}
        
% The following setting allows font size choices. Options are \tiny, \footnotesize, \small, \normalsize, \large, \Large, \LARGE, \huge, \Huge
\pgfplotsset{%
tick label style={font=\normalsize},
title style={font=\normalsize,align=center},
label style={font=\normalsize,align=center},
legend style={font=\normalsize}
            }

\begin{axis}[%
xlabel = {SNR (dB)},
ylabel = {MSE of the channel estimate},
%title  = {MSE comparison for the 3GPP Typical Urban channel scenario},
enlargelimits = true,
cycle list name = {color},
grid = major,
smooth,
scale = 1]

\addlegendentry{MLE}

\addplot [ color = {blue}, 
           mark  = {asterisk},
           style = {solid},
           line width = 1pt] table {\figDir/datFiles/fig1-line1.dat};

\addlegendentry{LMMSE}

\addplot [ color = {cyan}, 
           mark  = {square},
           style = {solid},
           line width = 1pt] table {\figDir/datFiles/fig1-line2.dat};

\addlegendentry{SURE Linear : $N = 3$}

\addplot [ color = {black}, 
           mark  = {o},
           style = {solid},
           line width = 1pt] table {\figDir/datFiles/fig1-line3.dat};

\addlegendentry{SURE Nonlinear : $N = 3$}

\addplot [ color = {magenta}, 
           mark  = {diamond},
           style = {solid},
           line width = 1pt] table {\figDir/datFiles/fig1-line4.dat};

\addlegendentry{CIR Thresholding \cite{Park2007}}

\addplot [ color = {red}, 
           mark  = {+},
           style = {solid},
           line width = 1pt] table {\figDir/datFiles/fig1-line5.dat};
       
\end{axis}
			\end{tikzpicture}
			\label{fig:mse3gpp}
		}

		\hspace{0.5cm}

		\subfigure[] %[Comparison of the bit error rate performance for the 3GPP typical urban channel scenario]
		{
			\begin{tikzpicture}
				\input{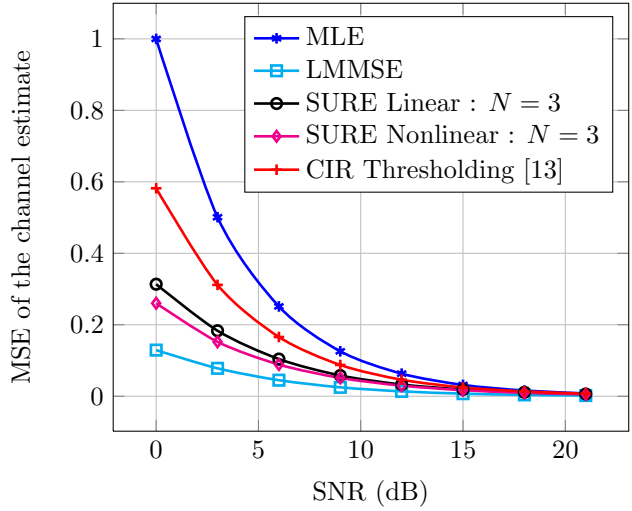}
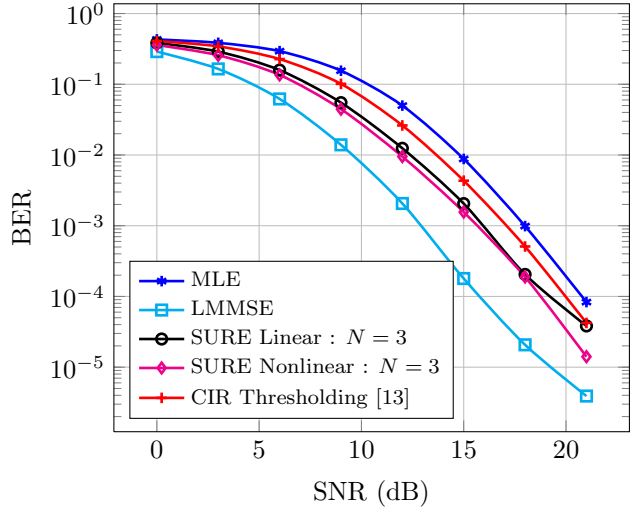
			\end{tikzpicture}
			\label{fig:ber3gpp}
		}
	}
\caption{Comparison of \subref{fig:mse3gpp} mean-square error and \subref{fig:ber3gpp} bit error rate performance of a receiver using LMMSE, ML, CIR-thresholding \cite{Park2007} and SURE-based (linear and nonlinear) channel estimates. The simulations has been performed with $ N = 3 $, OFDM symbol size $ K = 64 $, 16-QAM constellation, and for 3GPP typical urban channel scenario}
\label{fig:berCurvesVariousChannels}
\end{figure*}

In Figs. \ref{fig:mseCurvesVariousChannels} and \ref{fig:berCurvesVariousChannels}, we show the MSE and BER, at various values of SNR, for the ML, LMMSE, linear and nonlinear SURE-based estimators for various channel scenarios with OFDM symbol size $ K = 64 $. We see an improvement in the MSE and consequently, the BER for both SURE-based linear and nonlinear channel estimates. An interesting observation is that the performance of SURE-based nonlinear channel estimate is similar to that of its linear counterpart in 3GPP typical urban channel scenario, primarily because of the existance of few CIR taps with low SNR. However, for the single-tap Rayleigh channel environment, the nonlinear thresholding takes advantage of the presence of only one strong tap in the CIR, providing significant improvement with respect to SURE-based linear channel estimate. The performance improvement, in terms of SNR (dB) for $ 10^{-3} $, is tabulated in Table \ref{tab:sureMleTable}.

\begin{figure*}[t]
\centering
	{	
		\subfigure[] %[Comparison of the mean square error performance for single-tap Rayleigh channel]
		{	
			\begin{tikzpicture}
				% The following line determines the location of the legend
% E.g. - 1 : \pgfplotsset{every axis legend/.append style={legend pos = south west} , legend cell align = {left}}
% In E.g.2, the legend is placed at point (x,y) and the corner of the legend that sits on (x,y) is chosen by the anchor
% E.g. - 1 : \pgfplotsset{every axis legend/.append style={at={(0,0)},anchor=south west} , legend cell align = {left}}

\pgfplotsset{every axis legend/.append style={legend pos = north east},legend cell align = {left}}
        
% The following setting allows font size choices. Options are \tiny, \footnotesize, \small, \normalsize, \large, \Large, \LARGE, \huge, \Huge
\pgfplotsset{%
tick label style={font=\normalsize},
title style={font=\normalsize,align=center},
label style={font=\normalsize,align=center},
legend style={font=\footnotesize}
            }

\begin{axis}[%
xlabel = {SNR (dB)},
ylabel = {MSE of the channel estimate},
%title  = {MSE comparison for the single tap Rayleigh fading channel scenario},
enlargelimits = true,
cycle list name = {color},
grid = major,
smooth,
scale = 1]

\addlegendentry{MLE}

\addplot [ color = {blue}, 
           mark  = {asterisk},
           style = {solid},
           line width = 1pt] table {\figDir/datFiles/fig9-line1.dat};

\addlegendentry{LMMSE}

\addplot [ color = {cyan}, 
           mark  = {square},
           style = {solid},
           line width = 1pt] table {\figDir/datFiles/fig9-line2.dat};

\addlegendentry{SURE Nonlinear :  K = 64}

\addplot [ color = {magenta}, 
           mark  = {diamond},
           style = {solid},
           line width = 1pt] table {\figDir/datFiles/fig9-line3.dat};

%\addlegendentry{SURE Nonlinear : $N = 3$ : K = 128}
%
%\addplot [ color = {green}, 
%           mark  = {triangle},
%           style = {solid},
%           line width = 1pt] table {\figDir/datFiles/fig9-line4.dat};

\addlegendentry{SURE Nonlinear :  K = 256}

\addplot [ color = {green}, 
           mark  = {triangle},
           style = {solid},
           line width = 1pt] table {\figDir/datFiles/fig9-line5.dat};

%\addlegendentry{SURE Nonlinear : $N = 3$ : K = 512}
%
%\addplot [ color = {magenta}, 
%           mark  = {o},
%           style = {solid},
%           line width = 1pt] table {\figDir/datFiles/fig9-line6.dat};

\addlegendentry{SURE Nonlinear : K = 1024}

\addplot [ color = {brown}, 
           mark  = {x},
           style = {solid},
           line width = 1pt] table {\figDir/datFiles/fig9-line7.dat};

\end{axis}
			\end{tikzpicture}
			\label{fig:mseRayleighMultiFFT}			
		}

		\hspace{0.5cm}

		\subfigure[] %[Comparison of the bit error rate performance for the single-tap Rayleigh fading channel]
		{	
			\begin{tikzpicture}
				\input{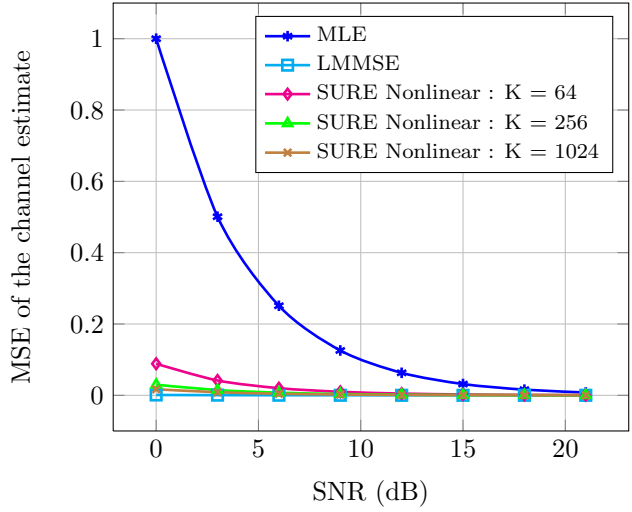}
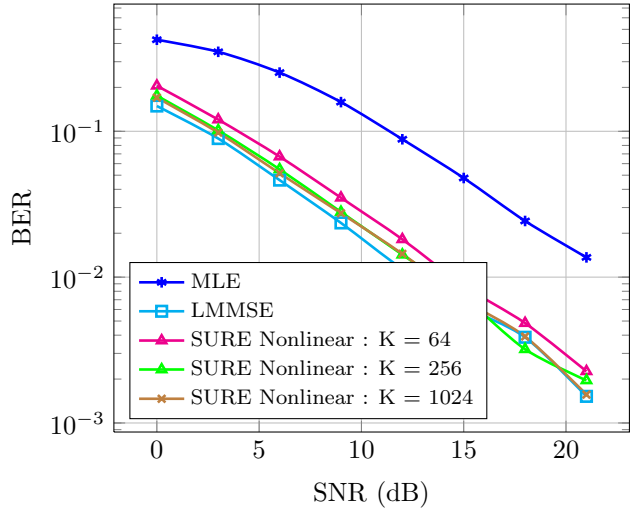
			\end{tikzpicture}
			\label{fig:berRayleighMultiFFT}			
		}		
	}
\caption{Comparison of the \subref{fig:mseRayleighMultiFFT} mean-square error and \subref{fig:berRayleighMultiFFT} bit error rate performance for the SURE-based (linear and nonlinear) channel estimates with $ N = 3 $, OFDM symbol size $ K = 64, 256 $ and $ 1024 $, and 16-QAM constellation. The simulation has been performed for the single-tap Rayleigh fading channel scenario}
\label{fig:mseFamilyOfCurves}
\end{figure*}

\begin{figure*}[t]
\centering
	{	
		\subfigure[] %[Comparison of the mean square error performance for the 3GPP typical urban channel]
		{
			\begin{tikzpicture}
				% The following line determines the location of the legend
% E.g. - 1 : \pgfplotsset{every axis legend/.append style={legend pos = south west} , legend cell align = {left}}
% In E.g.2, the legend is placed at point (x,y) and the corner of the legend that sits on (x,y) is chosen by the anchor
% E.g. - 1 : \pgfplotsset{every axis legend/.append style={at={(0,0)},anchor=south west} , legend cell align = {left}}

\pgfplotsset{every axis legend/.append style={legend pos = north east},legend cell align = {left}}
        
% The following setting allows font size choices. Options are \tiny, \footnotesize, \small, \normalsize, \large, \Large, \LARGE, \huge, \Huge
\pgfplotsset{%
tick label style={font=\normalsize},
title style={font=\normalsize,align=center},
label style={font=\normalsize,align=center},
legend style={font=\footnotesize}
            }

\begin{axis}[%
xlabel = {SNR (dB)},
ylabel = {MSE of the channel estimate},
%title  = {MSE comparison for the 3GPP Typical Urban channel scenario},
enlargelimits = true,
cycle list name = {color},
grid = major,
smooth,
scale = 1]

\addlegendentry{MLE}

\addplot [ color = {blue}, 
           mark  = {asterisk},
           style = {solid},
           line width = 1pt] table {\figDir/datFiles/fig7-line1.dat};

\addlegendentry{LMMSE}

\addplot [ color = {cyan}, 
           mark  = {square},
           style = {solid},
           line width = 1pt] table {\figDir/datFiles/fig7-line2.dat};

\addlegendentry{SURE Nonlinear : K = 64}

\addplot [ color = {magenta}, 
           mark  = {diamond},
           style = {solid},
           line width = 1pt] table {\figDir/datFiles/fig7-line3.dat};

%\addlegendentry{SURE Non-linear : $N = 3$ : K = 128}
%
%\addplot [ color = {green}, 
%           mark  = {triangle},
%           style = {solid},
%           line width = 1pt] table {\figDir/datFiles/fig7-line4.dat};

\addlegendentry{SURE Nonlinear : K = 256}

\addplot [ color = {green}, 
           mark  = {triangle},
           style = {solid},
           line width = 1pt] table {\figDir/datFiles/fig7-line5.dat};

%\addlegendentry{SURE Non-linear : $N = 3$ : K = 512}
%
%\addplot [ color = {magenta}, 
%           mark  = {o},
%           style = {solid},
%           line width = 1pt] table {\figDir/datFiles/fig7-line6.dat};

\addlegendentry{SURE Nonlinear : K = 1024}

\addplot [ color = {brown}, 
           mark  = {x},
           style = {solid},
           line width = 1pt] table {\figDir/datFiles/fig7-line7.dat};

\end{axis}
			\end{tikzpicture}
			\label{fig:mse3gppMultiFFT}								
		}
			
		\hspace{0.5cm}
		
		\subfigure[] %[Comparison of the bit error rate performance for the 3GPP typical urban channel scenario]
		{
			\begin{tikzpicture}
				\input{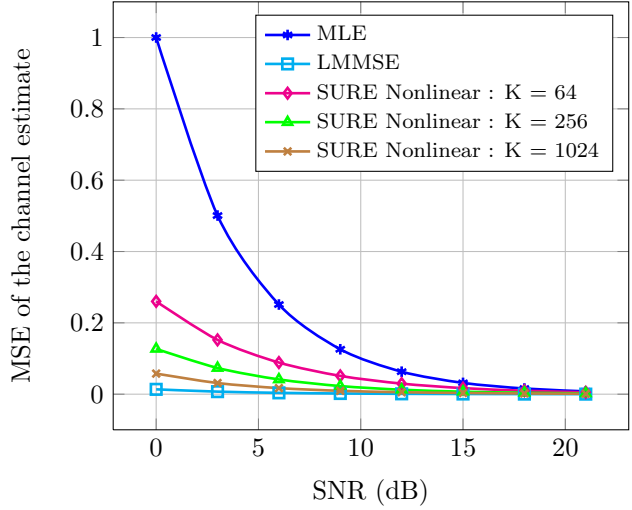}
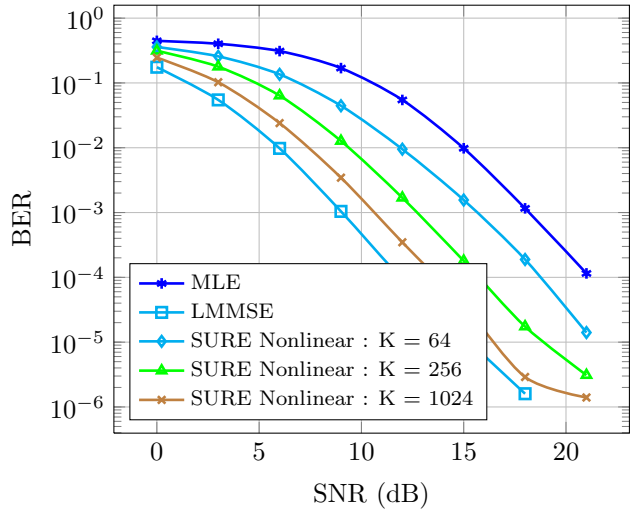
			\end{tikzpicture}
			\label{fig:ber3gppMultiFFT}					
		}
	}
\caption{Comparison of \subref{fig:mse3gppMultiFFT} the mean-square error and \subref{fig:ber3gppMultiFFT} bit error rate performance of a receiver using the SURE-based nonlinear channel estimate with $ N = 3 $, OFDM symbol size $ K = 64, 256 $ and $ 1024 $ and, 16-QAM constellation. The simulation has been performed for 3GPP typical urban channel scenario}
\label{fig:berFamilyOfCurves}
\end{figure*}

In Figs. \ref{fig:mseFamilyOfCurves} and \ref{fig:berFamilyOfCurves}, we present the MSE and BER, at various values of SNR, for the nonlinear SURE-based estimator for different values of OFDM symbol size $ K $ in various channel scenarios. The performance improvement can be attributed to two factors: (i) The variance of $ {\epsilon} $ is a function of the number of data points $ K $ used to estimate the mean-square error --- larger values of $ K $ reduce the variance of the risk estimate $ {\epsilon} $ and bring the parameters $ \mathbf{a}^\dagger $ closer to the MMSE solution; and (ii) Increasing the value of OFDM symbol size $ K $ for a channel with a given delay spread pads zeros to the CIR, which are taps with low SNR and can be attenuated for a better MSE.

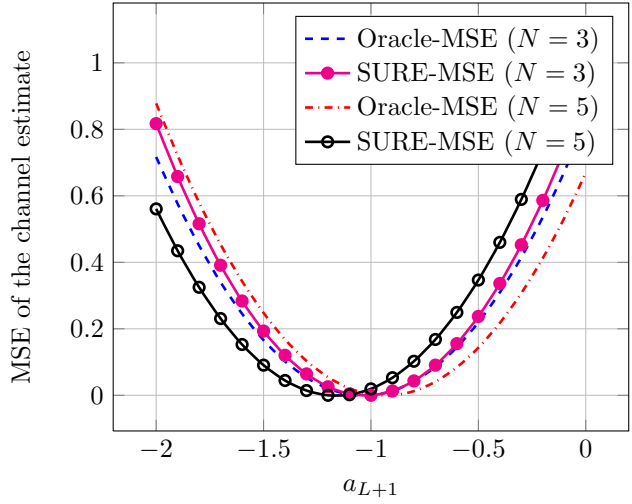
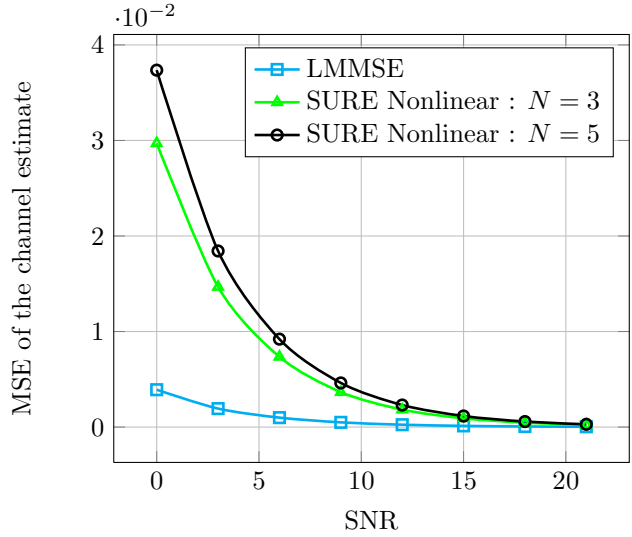
\begin{figure*}
\centering
	{	
		\subfigure[] %[Plot of the oracle MSE v/s $ {\epsilon} $ for different choices of the parameter $ a_{L+1} $]
			{
				\begin{tikzpicture}
					% The following line determines the location of the legend
% E.g. - 1 : \pgfplotsset{every axis legend/.append style={legend pos = south west} , legend cell align = {left}}
% In E.g.2, the legend is placed at point (x,y) and the corner of the legend that sits on (x,y) is chosen by the anchor
% E.g. - 1 : \pgfplotsset{every axis legend/.append style={at={(0,0)},anchor=south west} , legend cell align = {left}}

\pgfplotsset{every axis legend/.append style={legend pos = north east},legend cell align = {left}}
        
% The following setting allows font size choices. Options are \tiny, \footnotesize, \small, \normalsize, \large, \Large, \LARGE, \huge, \Huge
\pgfplotsset{%
tick label style={font=\normalsize},
title style={font=\normalsize,align=center},
label style={font=\normalsize,align=center},
legend style={font=\normalsize}
            }

\begin{axis}[%
xlabel = {$a_{L+1}$},
ylabel = {MSE of the channel estimate},
%title  = {Comparison of SURE estimate v/s the oracle value of the MSE},
enlargelimits = true,
cycle list name = {color},
grid = major,
smooth,
scale = 1]

\addlegendentry{Oracle-MSE $(N=3)$}

\addplot [ color = {blue}, 
           mark  = {none},
           style = {dashed},
           line width = 1pt] table {\figDir/datFiles/fig24-line1.dat};

\addlegendentry{SURE-MSE $(N=3)$}

\addplot [ color = {magenta}, 
           mark  = {*},
           style = {solid},
           line width = 1pt] table {\figDir/datFiles/fig24-line2.dat};

\addlegendentry{Oracle-MSE $(N=5)$}

\addplot [ color = {red}, 
           mark  = {none},
           style = {dashdotted},
           line width = 1pt] table {\figDir/datFiles/fig24-line4.dat};

\addlegendentry{SURE-MSE $(N=5)$}

\addplot [ color = {black}, 
           mark  = {o},
           style = {solid},
           line width = 1pt] table {\figDir/datFiles/fig24-line3.dat};

\end{axis}
				\end{tikzpicture}
				\label{fig:oracleVsSureMse}
			}
			
			\hspace{0.5cm}
		
		\subfigure[] %[Comparison of the mean square error performance for single-tap Rayleigh fading channel scenario]
		{	
			\begin{tikzpicture}
				% The following line determines the location of the legend
% E.g. - 1 : \pgfplotsset{every axis legend/.append style={legend pos = south west} , legend cell align = {left}}
% In E.g.2, the legend is placed at point (x,y) and the corner of the legend that sits on (x,y) is chosen by the anchor
% E.g. - 1 : \pgfplotsset{every axis legend/.append style={at={(0,0)},anchor=south west} , legend cell align = {left}}

\pgfplotsset{every axis legend/.append style={legend pos = north east},legend cell align = {left}}
        
% The following setting allows font size choices. Options are \tiny, \footnotesize, \small, \normalsize, \large, \Large, \LARGE, \huge, \Huge
\pgfplotsset{%
tick label style={font=\normalsize},
title style={font=\normalsize,align=center},
label style={font=\normalsize,align=center},
legend style={font=\normalsize}
            }

\begin{axis}[%
xlabel = {SNR},
ylabel = {MSE of the channel estimate},
%title  = {MSE Comparison for the 3GPP Typical Urban channel scenario},
enlargelimits = true,
cycle list name = {color},
grid = major,
smooth,
scale = 1]

\addlegendentry{LMMSE}

\addplot [ color = {cyan}, 
           mark  = {square},
           style = {solid},
           line width = 1pt] table {\figDir/datFiles/fig13-line1.dat};

\addlegendentry{SURE Nonlinear : $N = 3$}

\addplot [ color = {green}, 
           mark  = {triangle},
           style = {solid},
           line width = 1pt] table {\figDir/datFiles/fig13-line2.dat};

\addlegendentry{SURE Nonlinear : $N = 5$}

\addplot [ color = {black}, 
           mark  = {o},
           style = {solid},
           line width = 1pt] table {\figDir/datFiles/fig13-line3.dat};

\end{axis}
			\end{tikzpicture}
			\label{fig:sureOverparameterizedMse}			
		}
	}
\caption{\subref{fig:oracleVsSureMse} Plot of the oracle MSE v/s $ {\epsilon} $ for different choices of the parameter $ a_{L+1} $. \subref{fig:sureOverparameterizedMse} Comparison of the mean-square error performance for single-tap Rayleigh fading channel scenario. The simulation has been performed for the single-tap Rayleigh fading channel scenario with $ 0 $ dB SNR and OFDM symbol size $ K = 256 $}	

\end{figure*}

In Fig. \ref{fig:oracleVsSureMse}, we compare the oracle MSE with Stein's unbiased estimate of the risk as a function of the parameter $ a_{L+1} $. We observe that, for $ N = 3 $, the minima of both curves are close to each other, indicating that the value of $ a_{L+1} $ obtained by minimizing SURE is close to the value obtained by minimizing the oracle MSE. However, this is not the case for $ N = 5 $ since the variance of SURE increases with the number of parameters that have to be estimated. In Fig. \ref{fig:sureOverparameterizedMse}, we observe that SURE-based channel estimate with $ N = 5 $ has a higher MSE than that with $ N = 3 $. Therefore, from our simulations, we conclude that for practically used values of OFDM symbol sizes, $ N = 3 $ provides the maximum performance improvement.

%\begin{table}[ht]
%\begin{tabularx}{\linewidth}{XXXX}
%\hline
%Channel type & AWGN & Single-tap Rayleigh Channel ($ 10^{-2} $ BER) & 3GPP Typical Urban Channel \\
%\hline
%$ N = 1 $	& 0.31 dB & 0.16 dB & 0.15 dB \\
%\hline
%$ N = 3 $	& 3.65 dB & 3.63 dB & 1.97 dB \\
%\hline
%$ N = 5 $	& 4.37 dB & 4.63 dB & 2.21 dB \\
%\hline
%$ N = 7 $	& 4.43 dB & 4.67 dB & 2.37 dB \\
%\hline
%\end{tabularx}
%\caption{Performance improvement in SNR (dB) for SURE optimized non-linear channel estimate over the ML estimate for different channels.}
%\label{tab:sureMleTable}
%\end{table}

\section{Conclusions}
\label{section:conclusion}
We considered the preamble-based channel estimation problem in an OFDM system. Modelling the CFR as a vector of unknown deterministic parameters, we proposed a method based on minimizing SURE to derive optimal parameters of denoising functions to obtain channel estimates. We showed that SURE-optimized channel estimation algorithm provided a significant improvement over the conventional ML estimate of the channel in both the linear and nonlinear formulations. For channels with a large number of low SNR taps, the performance of the non-linear SURE based function, while not assuming prior knowledge of the channel statistics, was shown comparable to the LMMSE channel estimate. However, the performance improvement was dependent on the variance of $ {\epsilon} $, which from simulation, was found to be acceptable for $ N = 3 $ and for values of OFDM symbol size $ K $ and SNR commonly encountered in practice. Moreover, The algorithm for $ N = 3 $ requires a $ 4 \times 4 $ matrix inversion and has the advantage of practical implementability coupled with a performance improvement of $ 2.25 $ dB in realistic channel scenarios. 

%We also see that there is a performance improvement when we increase the number of carriers we use to de-noise each subcarrier ($ N $).  This however has an upper-limit for two reasons; One of them being, larger number of parameters lead to a larger variance in the estimate of the MSE making the minimization of this risk very susceptible to noise; The other reason being, practical implementation restricts us from inverting large matrices (beyond $ N = 3 $). However, we see that there is atleast a $ 2.25 $ dB gain in SNR with $ N = 3 $ (for the 3GPP channel) rendering this method worthwhile for practical implementation.

%Since larger values of $ K $ leads to a lower variance of SURE, we need to keep in mind that the choice of $ N $ depends on the value of $ K $.

%\input{SetA1/apx.tex}        
%\bibliographystyle{IEEEtran}
%\bibliography{IEEEabrv,paperBibliography.bib}

\bibliographystyle{elsarticle-num} 
\bibliography{paperBibliography}

\newpage
\clearpage
\renewcommand{\arraystretch}{1.4}% Tighter
\begin{table}[t]
\begin{tabularx}{\linewidth}{XXXX}
\hline
Channel type & AWGN & Single-tap Rayleigh channel ($ 10^{-1.5} $ BER) & 3GPP typical urban channel \\
\hline 
$ N = 3 $	& 6.88 dB & 7.26 dB & 2.35 dB \\
\hline
\end{tabularx}
\vspace{0.5cm}
\caption{Performance improvement in SNR (decibel) for SURE-optimized nonlinear channel estimate over the ML estimate for different channels and $ K = 64 $.}
\label{tab:sureMleTable}
\end{table}

\newpage
\clearpage

\newpage
\clearpage

\newpage

\end{document}